\documentclass[preprint,preprintnumbers,nofootinbib,superscriptaddress]{revtex4}
\usepackage{amsmath}
\usepackage{amssymb}
\usepackage{graphicx}
\usepackage{subfigure}
\newcommand{\nc}{\newcommand}

\nc\beq{\begin{equation}}
\nc\eeq{\end{equation}}
\nc\beqa{\begin{eqnarray}}
\nc\eeqa{\end{eqnarray}}
\nc\bi{\begin{itemize}}
\nc\ei{\end{itemize}}
\def\nn{\nonumber \\ }

\long\def\sl#1{\hbox{\tiny #1}}			
\long\def\teq#1{\hbox{$#1$}}				

\nc{\vv}{\boldsymbol}					
\nc{\La}{{\cal L}}

\nc{\bert}{\raise-0.45mm\hbox{\Large$\Box$}}	
\nc\A{A^\mu}
\nc{\PgP}{\bar\psi\gamma^\mu\psi}			
\nc{\tensorbilinear}{\bar\psi\frac{i}{2}(\gamma_\mu
\buildrel\rightarrow\over\partial_\nu -
\gamma_\mu\buildrel\leftarrow\over\partial_\nu)\psi}

\renewcommand{\slash}[1]{#1\!\!\!/}

\def\Dsl{\,\raise.15ex \hbox{/}\mkern-12.8mu D}
\newcommand\Tr{{\rm Tr\,}}

\def\OMIT#1{}

\newcommand{\ba}{\begin{array}}
\newcommand{\ea}{\end{array}}

\DeclareMathOperator{\Real}{Re}
\DeclareMathOperator{\Imag}{Im}

\begin{document}
\preprint{MIT-CTP-3834}
\preprint{CALT-68-2645}

\title{Signs of Analyticity in Fermion Scattering}
\author{Allan~Adams}
\author{Alejandro~Jenkins}
\affiliation{Center for Theoretical Physics, Laboratory for Nuclear Science and Department of Physics, Massachusetts Institute of Technology, Cambridge, MA 02139, USA}
\author{Donal~O'Connell}
\affiliation{School of Natural Sciences, Institute for Advanced Study, Princeton, NJ 08540, USA}
\affiliation{The Niels Bohr Institute, Blegdamsvej 17, DK-2100 Copenhagen, Denmark}
\affiliation{Department of Physics, California Institute of Technology, Pasadena, CA 91125, USA}

\begin{abstract}
\medskip
\noindent

We show that the signs of the leading irrelevant interactions for Dirac fermions are constrained by the analytic structure of the $S$-matrix.  If Regge behavior obtains, negative signs indicate the presence of higher-spin bound states that spoil the convergence of the dispersion integrals and drive the corresponding operators relevant.  For nucleon-nucleon scattering, the negativity of some of the low-energy interactions signals the presence of a spin-1 bound state: the deuteron.  We connect the divergence of the dispersion integral to the ``Sommerfeld enhancement'' of the cross-section for low-energy scattering.  We also discuss how this illuminates potential pitfalls in using perturbative methods to understand the dependence of the low-energy nuclear interaction on the masses of the light quarks.  Finally, we suggest the possibility of applying similar reasoning to the current-current operators in the electroweak effective lagrangian, where no bound states spoil convergence of the dispersion relations.

\end{abstract}

\maketitle

\newpage

\section{Introduction}
\label{intro}

It is widely believed that locality and microcausality are encoded in the analytic structure of the $S$-matrix, with physical scattering amplitudes arising as real boundary values of functions which are analytic up to poles and branch cuts dictated by unitarity \cite{microcausality}.  The constraints implied by this analyticity were the subject of intense study in the 1960's and 70's.\footnote{For in-depth textbook treatments, see, for example, \cite{ELOP}.}  More recently, such constraints have been translated into effective field theory, where they appear as bounds on the coefficients of leading irrelevant operators (precisely the ones which must be UV completed) such as those in the chiral lagrangian \cite{chiralpositivity,ANSAR,jacques}.  That these bounds also prevent macroscopic violations of locality and causality \cite{ANSAR} and ensure the stability of auxiliary intermediates \cite{O'Jenkins} highlights the role of analyticity in ensuring these physical properties.

In this article we deduce bounds on low-energy contact interactions of fermions which similarly derive from the analyticity of the $S$-matrix.  As an example of the theories we will constrain, consider the effective description of a single massive fermion,
\beq
\label{EQ.Example}
\La = \bar \psi \left( i \slash \partial - m \right) \psi + \frac{a}{M^{2}} \left(\bar \psi \psi \right)^2 + \ldots
\eeq
where $M$ is the effective cutoff and $a$ is a classically dimensionless coupling.  This theory is manifestly non-renormalisable.  Given a UV completion, however, the effective theory gives a consistent expansion for its $S$-matrix in powers of $s$ which is unitary and analytic at scales below the cutoff.  But when does a UV completion exist?   Following \cite{O'Jenkins}, suppose we generated the operator $(\bar\psi \psi)^{2}$ by integrating out an auxiliary intermediate from the action,
\beq
\La' = \bar \psi \left( i \slash \partial - m \right) \psi + F \left(\bar \psi \psi \right)  - \frac{M^{2}}{4 a} F^{2}~.
\label{dP}
\eeq
Performing the (trivial) path integral over $F$ returns the original lagrangian of Eq. (\ref{EQ.Example}).  However, if $a<0$, the path integral diverges, and if $F$ were made dynamical then its potential would drive it to a non-zero VEV.  Perturbations about a stable minimum of the potential would have a strictly positive effective coupling, $a_{\sl{eff}}>0$.  This suggests that the original effective field theory of $\psi$ requires both UV {\em and} IR completion when \teq{a < 0}.  The observation that one sign of an effective 4-fermion interaction cannot be obtained from integrating out a heavy, non-tachyonic mediator and is therefore ``unnatural'' has also been presented in \cite{hashimoto}.

In what follows we will study the constraints on effective four-fermion operators by considering the dispersion relations which must be satisfied by any UV completion of Eq. (\ref{EQ.Example}) with an analytic and unitary $S$-matrix.  A basic object in these arguments is the spectral density over cuts in the complex $t$-plane.  When this density has localized support, it can be approximated by a resonance whose width is controlled by the spectral density.  This is the resonance represented by $F$ in the toy model of Eq. (\ref{dP}).

All of this assumes that the dispersion integral of interest converges.  If the spectral function does {\em not} vanish asymptotically as $t \to \infty$, then a narrow resonance approximation is of course impossible.  The dispersion integral diverges and must be subtracted, introducing additional terms to the dispersion relation which leave the sign of the contact interaction undetermined.   If Regge theory applies, such a divergence can be traced to the presence of two-fermion composite states with spin 1 or higher: a violation of the naive sign constraint thus suggests the existence of bound states with spin.  We shall see this in detail in the case of nucleon-nucleon scattering, where the existence of the deuteron bound state is associated with the violation of these sign constraints.

\section{Invariant amplitudes}
\label{invariants}

The analytic structure of fermion-fermion scattering was first explored in \cite{nambu}.  Let $\psi$ be a Dirac fermion with mass $m$.  The amplitude for relativistic \teq{\psi \psi \to \psi \psi} scattering may be expressed as
\beq
{\cal M}(p_1, p_2 \to p_1', p_2') =  \bar u^{s_1'}(p_1') \bar u^{s_2'}(p_2')
\left( \sum_i c_i(s,t) \hat C_i \right) u^{s_1}(p_1) u^{s_2}(p_2)~,
\label{fermionM}
\eeq
where the invariant amplitudes $c_i$ are scalar functions of the Mandelstam variables \teq{s = (p_1 + p_2 )^2}, \teq{t = (p_1 - p_1' )^2} and \teq{u = (p_1 - p_2' )^2}.  Since \teq{s+t+u = 4 m^2}, we may write the $c_{i}$ as functions of $s$ and $t$ alone.  The $\hat C_i$'s are matrices in spinor space.   When the interaction is parity-conserving and time-translation invariant, we need only five $\hat C_i$'s, which we may write as
\beq
\hat C_i = \eta_i \, \Gamma_i^{(1)} \cdot \Gamma_i^{(2)}~,
\label{Gamma}
\eeq
where $\eta_i$ is some arbitrary numerical factor, while the dot represents the appropriate contraction of Lorentz indices, and the superscript indicates whether the spin matrix $\Gamma_i$ acts between \teq{\bar u^{s_1'}(p_1')} and \teq{u^{s_1}(p_1)}, or between \teq{\bar u^{s_2'}(p_2')} and \teq{u^{s_2}(p_2)}, respectively.  The {\em exchange terms} in the amplitude, arising from the matrix acting between \teq{\bar u^{s_2'}(p_2')} and \teq{u^{s_1}(p_1)}, or between \teq{\bar u^{s_1'}(p_1')} and \teq{u^{s_2}(p_2)}, may be absorbed into the $c_i$'s of Eq. (\ref{fermionM}) through Fierz transformations \cite{goldberger}.

A convenient basis is
\beq
\Gamma_1 = 1;~ \Gamma_2 = \gamma^\mu;~ \Gamma_3 = i \gamma_5;~ \Gamma_4 = \gamma^\mu \gamma_5;~ \Gamma_5 = \sigma^{\mu \nu}~.
\eeq
where \teq{\sigma^{\mu\nu} \equiv i \left[ \gamma^\mu, \gamma^\nu \right]/2}.  Notice that \teq{ \gamma^0 \Gamma_i^\dagger \gamma^0 = \Gamma_i}, which makes all \teq{\left(\bar \psi \Gamma_i \psi \right)}'s hermitian.  For reasons that will become clear later, when working in the $(+---)$ metric convention favored by particle physicists, we choose to define
\beqa
\hat C_1 = 1^{(1)} 1^{(2)}; ~\hat C_2 = -\left(\gamma^{\mu} \right)^{(1)} \left(\gamma_\mu\right)^{(2)};~
\hat C_3 = (i \gamma_5)^{(1)} (i \gamma_5)^{(2)};~ \nn
\hat C_4 = -\left(\gamma^\mu \gamma_5 \right)^{(1)} \left( \gamma_\mu  \gamma_5 \right)^{(2)};~
\hat C_5 = \frac 1 2 \left( \sigma^{\mu \nu} \right)^{(1)} \left( \sigma_{\mu \nu} \right)^{(2)}~.
\label{fermionC}
\eeqa
(If we were working in the $(-+++)$ convention, we would omit the minus signs in front of $\hat C_2$ and $\hat C_4$.)

\section{Fixed-$s$ dispersion relation}
\label{dispersion}

\def\ct{{\tilde c}}

For fixed $s$, the $c_i(s,t)$'s must be analytic functions of complex $t$, except for possible poles and branch cuts along the real axis.  By Cauchy's theorem, if the $c_i(s,t)$'s go to zero as $t \to \infty$, they must obey the fixed-$s$, unsubtracted dispersion relation
\beq
c_i(s,t,u) = \frac 1 {2 \pi i} \int_0^\infty dt' \, \frac{D_i^{(t)}(s,t',u')}{t'-t}
+ \frac 1 {2 \pi i} \int_0^\infty du' \frac{D_i^{(u)}(s,t',u')}{u'-u}~,
\label{dispfull}
\eeq
with $u' = 4m^2 - s - t'$.  $D_i^{(t)}$ is the $t$-channel discontinuity, which for fixed $s=0$ consists of pole contributions of the form \teq{2\pi i g^2 \delta(t - M^2)} for exactly stable mediator particles with mass $M$ and coupling $g$, and a branch cut discontinuity $2 i \Imag c_i(s, t)$ for $t > 4 m_{\sl{gap}}^2$, where $m_{\sl{gap}}$ is the mass of the lightest particle into which a mediator may decay.  The $u$ channel discontinuity \teq{D_i^{(u)}} corresponds to poles and cuts in the {\em left}-hand side of the complex $t$-plane, with \teq{t < 4(m^2 - m_{\sl{gap}}^2)}.

It is useful to define $\ct_i$'s corresponding only to the {\it direct terms} in the scattering amplitude.  By Fermi statistics, we may express the amplitude in Eq. (\ref{fermionM}) as
\beq
{\cal M} = \bar u^{s_1'}(p_1') \bar u^{s_2'}(p_2')
\left( \sum_i \ct_i(s,t,u) \hat C_i - \sum_{i,j} \ct_i (s,u,t) F_{ij} \hat C_j \right)
u^{s_1}(p_1) u^{s_2}(p_2)
\label{ctilde}
\eeq
where $F$ is the matrix defined by the Fierz identitites, which in the basis of Eq. (\ref{fermionC}) takes the form
\beq
F_{ij} = \frac 1 4 \left( \begin{array}{r r r r r}
1 & -1 & -1 & 1 & 1 \\ 
- 4 & -2 & -4 & -2 & 0 \\
-1 & -1 & 1 & 1 & -1 \\
4 & -2 & 4 & -2 & 0 \\
6 & 0 & -6 & 0 & -2 \\
\end{array} \right)~.
\label{fierz}
\eeq
We may then consider the simplified dispersion relation
\beq
\ct_i(s,t) = \frac 1 \pi \int_0^\infty dt' \, \frac{\Imag c_i(s,t')}{t'-t}
\label{disp}
\eeq
over the positive real axis only.\footnote{The dispersion relation of Eq. (\ref{disp}) was introduced for the study of the correlated two-pion exchange contribution to the nucleon-nucleon scattering by Cottingham, Vinh Mau, and others \cite{cottingham}.  For more modern textbook treatments, see \cite{textbooks}.  The form of this dispersion relation is also reviewed clearly in \cite{serot}.}

Another advantage of the $\ct_i$'s is that, when the mass of the scattered particles is large compared to the other energy scales of interest,
\beq
\left| \ct_i \left(0, 0, 4m^2 \right) \right| \gg \left| \ct_i \left(0, 4m^2, 0 \right) \right|~.
\label{nonrel1}
\eeq
For instance, if the only interaction were the exchange of a stable scalar boson with mass $\mu$ and coupling $g$, we would have \teq{|\ct_1 (0,0,4m^2)| = g^2 / \mu^2 \gg |\ct_1(0,4m^2,0)| = g^2/(4m^2 - \mu^2)}.  In general, for fixed $s=0$ in the non-relativistic limit,
\beq
c_i (0, 0) \approx \ct_i(0,0)~,
\label{nonrel2}
\eeq
so that the exchange term in Eq. \ref{ctilde} can be neglected \cite{serot}, leaving only the $t$-channel cut for the $c_i$'s in Eq. (\ref{dispfull}).

\begin{figure}[t]


\includegraphics[bb=150 370 470 550,clip,scale=1]{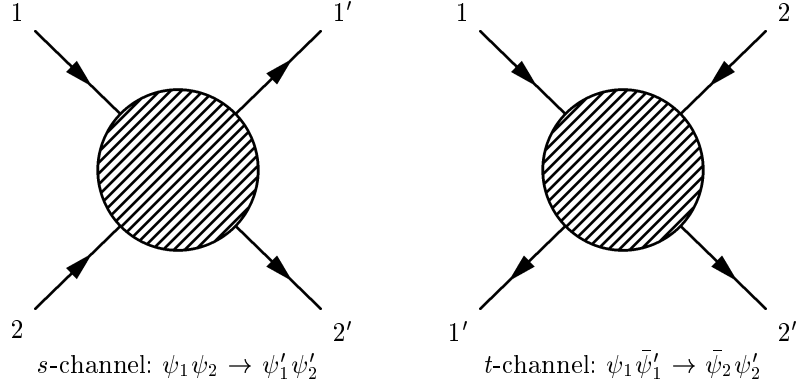}

\caption{Diagrams representing the $s$ channel and $t$ channel scattering processes respectively, related by crossing of the particles labeled $1'$ and $2$.  (In this paper, Feynman diagrams are always drawn with time flowing from left to right.)}
\label{channels}
\end{figure}

We may extract the value of $\Imag c_i (s,t)$ for unphysical $t>0$ from the $t$-channel \teq{\psi \bar \psi \to \bar \psi \psi} scattering process obtained by crossing $p_2$ with $p_1'$ (see Fig. \ref{channels}).  In this channel the roles of $s$ and $t$ are exchanged.  We may write the corresponding amplitude as
\beq
{\cal M} \left(p_1, \bar p_1' \to \bar p_2, p_2' \right)  = \bar v^{\bar s_1'}(p_1') \bar u^{s_2'}(p_2')
\left( \sum_i c_i(t,s) \hat C_i \right) u^{s_1}(p_1) v^{\bar s_2}(p_2)~.
\label{tchannel}
\eeq

Unitarity of the $S$-matrix requires that
\beqa
-i \left[ {\cal M} \left(p_1, \bar p_1' \to \bar p_2, p_2' \right) - {\cal M}^\ast \left(\bar p_2, p_2' \to p_1, \bar p_1'\right) \right] = \nn
\sum_n \rho_n \, {\cal M}^\ast \left(\bar p_2, p_2' \to n \right) {\cal M} \left(p_1, \bar p_1' \to n \right)
\label{unitarity}
\eeqa
where the summation is over all allowed channels and $\rho_n$ is the phase space factor (energy-momentum conserving delta functions have been suppressed).  The right-hand side of Eq. (\ref{unitarity}) can be expanded in the form
\beq
\bar v^{\bar s_1'}(p_1') \bar u^{s_2'}(p_2')
\left[ \sum_i d_i(t,s) \hat C_i \right] u^{s_1}(p_1) v^{\bar s_2}(p_2)~.
\label{di}
\eeq
By choosing different helicities for the incoming and outgoing particles, Eq. (\ref{unitarity}) generates five independent constraints,\footnote{See \cite{goldberger} for the proof that the scattering can be characterized by five independent helicity amplitudes, and for the transformation between the helicity amplitudes and the invariant $c_i$'s in $s$-channel scattering.  For the transformation in the $t$-channel see, e.g., the appendix in \cite{serot}.} which may be re-expressed in terms of the invariant amplitudes in the simple form
\beq
\Imag c_i (t, s) = d_i (t,s)~.
\eeq
We expect $d_i(t,s=0)$ to be non-negative, because in that case \teq{p_1 = p_2} and \teq{p_1' = p_2'}, making the two ${\cal M}$'s in each term of the sum on the right-hand side of Eq. (\ref{unitarity}) equivalent up to the assignment of external spinors, which are removed to construct the $d_i$'s.

For further guidance, let us consider elastic unitarity, where the sum over channels in Eq. (\ref{unitarity}) is restricted to fermion-antifermion states.  If we express the spinor matrices \teq{\hat C_i} in the form of Eq. (\ref{Gamma}) then, after summing over the spins and momenta of the intermediate fermion and antifermion, we find that:
\beqa
d_i (t, s = 0) &=& \eta_i
\sum_j \int \frac{d^3 k_1}{(2\pi)^3}\frac 1 {2E_1} \frac{d^3 k_2}{(2\pi)^3}\frac 1 {2E_2}
c_i(t, w) c^\ast_j (t, w) \, \Tr\left[ \Gamma_i (\slash k_1 + m) \Gamma_j (\slash k_2 - m ) \right] \nn
&& \hskip 45pt \times (2 \pi)^4 \delta^{(4)} (p_1 + p_1' - k_1 - k_2)
\label{elastic}
\eeqa   
where \teq{t = (p_1 + p_1')^2 \geq 4 m^2}, \teq{w \equiv (p_1 - k_2)^2}, and \teq{k_1^2 =  k_2^2 = m^2}.  By the Dirac equation
\beq
\bar v(p_1') ( \slash k_1 + \slash k_2 )  u(p_1) = \bar v(p_1') ( \slash p_1 + \slash p_1' )  u(p_1) = 0~.
\eeq
It is then a straightforward though somewhat tedious exercise in gamma-matrix algebra to show that there is no contribution to $d_i$ in Eq. (\ref{elastic})) for $j \neq i$ and that the $d_i(t,s=0)$'s are all non-negative.\footnote{The Lorentz indices on $\Gamma_{i,j}$ in Eq. (\ref{elastic}) are not contracted with each other, but rather with the indices of the corresponding \teq{\Gamma}'s inside the $\hat C_i$'s of Eq. (\ref{di}).  The factors of $\eta_i$ in Eq. (\ref{fermionC}) were chosen to match the sign-definiteness of the trace factor in Eq. (\ref{elastic}), so that the \teq{d_i(t,0)}'s would all come out non-negative.  Since the sign of the trace depends on the metric convention, so does the convenient choice of $\eta_i$'s.}

\begin{figure}[t]



%
%
%
%
%
%
%
%

\includegraphics[scale=1]{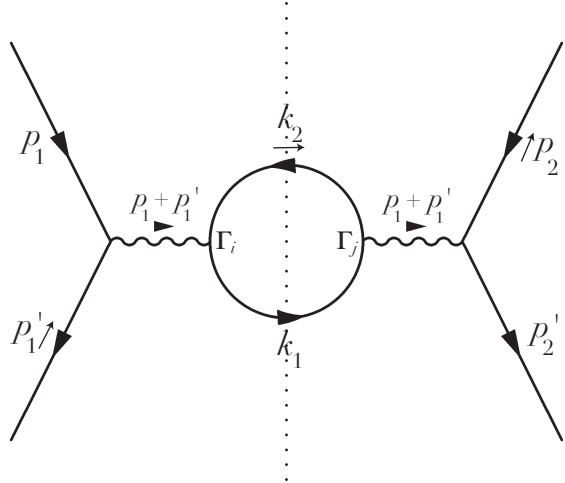}

\caption{Elastic unitarity can be visualized as the cutting across two internal fermion lines, yielding Eq. (\ref{elastic}).}
\label{cutkosky}
\end{figure}

In terms of the Cutkosky rules, elastic unitarity corresponds to cutting diagrams only across two fermion lines, as shown in Fig. \ref{cutkosky}.  Whatever lines we cut across, though, they must correspond to a state with the correct quantum numbers to be produced on-shell from the fermion-antifermion interaction.  After summing over all intermediate spins and momenta, the factor resulting from the cutting must, like the trace in Eq. (\ref{elastic}), be diagonal in $i$ and $j$, and it seems very plausible that they must have the same sign definiteness, due to the form of the right-hand side of Eq. (\ref{unitarity}) that is dictated by unitarity.

Therefore, if we consider the contributions to each of the invariant amplitudes $c_i (t, s=0)$'s diagram by diagram, applying the cutting rules to each, we conclude that
\beq
\Imag \teq{c_i (s=0, t > 0) \geq 0}~.
\label{positivity}
\eeq
It then follows from Eq. (\ref{disp}) that the $\ct_i(0,0)$'s must be non-negative in any theory with an analytic $S$-matrix for which \teq{\ct_i(0,t) \to 0} as \teq{t \to \infty}.

\section{Constraints on the effective action}
\label{zeroenergy}

For a single Dirac fermion with all mediators integrated out ---and assuming parity and time-translation invariance, as well as Lorentz invariance--- we may express the interaction part of the full quantum-mechanical effective action $\Gamma_{\sl{1PI}}$, in terms of the $\tilde c_i$'s defined in Eq. (\ref{ctilde}), as:
\beqa
\int \prod_{j=1}^4 d^4 p_j \, (2 \pi)^4 \delta^{(4)}\left(p_1 + p_2 - p_3 - p_4 \right)
\, \sum_{i=1}^5 \, \ct_i \left( (p_1 + p_2)^2, (p_1 - p_3)^2, (p_1 - p_4)^2 \right) \nn
\times \left[\bar \psi (p_4) \Gamma_i \psi (p_2) \right] \cdot \left[\bar \psi (p_3) \Gamma_i \psi (p_1) \right]
~,
\label{Gamma1PI}
\eeqa
where the fields $\psi$ are written in momentum-space.  In this language, it is clear how the direct and exchange terms in the scattering amplitude come from the two distinct contractions of the interaction operators with the two-particle in and out states.  It is also clear why, e.g., the values of the $\ct_i(s, t)$'s for $t > 4m^2$ and $s < 0$ can be extracted from particle-antiparticle scattering, as illustrated in Fig. \ref{channels}.

Expanding Eq. (\ref{Gamma1PI}) around zero energy we can write an effective Lagrangian
\beqa
\La_{\sl{eff}} &=&  \bar \psi \left( i \slash \partial - m \right) \psi + \sum_{i=1}^5 a_i \, \eta_i \left( \bar \psi \Gamma_i \psi \right)^2 + \ldots \nn
&=& \bar \psi \left( i \slash \partial - m \right) \psi + a_1 \left(\bar \psi \psi \right)^2 - a_2 \left(\bar \psi \gamma ^\mu \psi \right)^2 + a_3 \left(i \bar \psi \gamma_5 \psi \right)^2 \nn
& & - a_4 \left(\bar \psi \gamma ^\mu \gamma_5 \psi \right)^2 + \frac{a_5}{2} \left(\bar \psi \sigma^{\mu \nu} \psi \right)^2 + \ldots~,
\label{effective}
\eeqa
with \teq{a_i = {\ct_i(0,0,0)}}.  Our positivity argument from Sec. \ref{dispersion} therefore translates into constraints \teq{a_i \geq 0} on the coefficients of the interactions for effective theories of the form Eq. (\ref{effective}) that admit analytic UV completions.\footnote{If instead of the full quantum-mechanical $\Gamma_{\sl{1PI}}$, we consider an ordinary perturbative low-energy effective action, then there might be some renormalization-scheme dependence in translating from \teq{\ct_i(0,0,0)} to the coefficient $a_i$.}  Notice that these constraints agree with the argument sketched in \cite{O'Jenkins} that the analytic constraints on possible low-energy interactions are equivalent to forbidding tachyons in a narrow-resonance UV completion.

In formulating our constraints in terms of the coefficients of effective operators, one must bear in mind, however, that the action in Eq. (\ref{effective}) can always be rewritten by applying the Fierz identities to any part of any of the interaction terms.  The resulting action may look very different, and some of the $a_i$'s may change signs, but it will predict identical zero-energy scattering amplitudes.

To remedy this ambiguity, we diagonalize the matrix $F$ in Eq. (\ref{fierz}), which has five linearly independent eigenvectors: three with eigenvalue -1 and two with eigenvalue 1.  Thus there are three independent 4-fermion operators that transform unto themselves under fierzing, which we may write in the form
\beq
{\cal O}_i = \eta_i \left( \bar \psi \Gamma_i \psi \right)^2 - \sum_j F_{ij} \eta_j \left( \bar \psi \Gamma_j \psi \right)^2
\label{goodF}
\eeq
for some choice of $i$.  The coefficients of these three independent operators in the zero-energy effective action must be positive by the argument presented in Sec. \ref{dispersion}.

The two operators that fierz into minus themselves, corresponding to the eigenvalues 1 of $F$ and expressible in the form
\beq
{\cal O}_i = \eta_i \left( \bar \psi \Gamma_i \psi \right)^2 + \sum_j F_{ij} \eta_j \left( \bar \psi \Gamma_j \psi \right)^2
\label{badF}
\eeq
give no interaction at zero energies: in Eq. (\ref{ctilde}) the exchange term exactly cancels the direct term for the zeroth-order term in the power expansion of the scattering amplitude as a function of the Mandelstam variables $s,t,u$.  Therefore there are no constraints on their coefficients in the zero-energy effective action:  Adding any linear combination of the operators of the form Eq. (\ref{badF}) leaves the zero-energy amplitudes unchanged while destroying the simple relations between the $\ct_i(0,0,0)$'s constrained in Sec. \ref{dispersion} and the $a_i$'s of Eq. (\ref{effective}).

Notice that the $a_i$'s in Eq. (\ref{effective}) describe interactions with $s=t=u=0$, which is not a good limit nonrelativistically because it corresponds to massless particles (or particles whose masses are negligible compared to the energy scales of interest).  Non-relativistically, we may neglect the exchange term in Eq. (\ref{ctilde}) and treat the particles as distinguishable, as we argued in Sec. \ref{dispersion}.  We may therefore obtain the analytic constraints on the nonrelativistic effective action simply by carrying out the non-relativistic expansion of Eq. (\ref{effective}) with $a_i \geq 0$.

\section{Asymptotic behavior and nucleon scattering}
\label{asymptotic}

Our constraints do not necessarily apply if the dispersion integrals do not converge, i.e., if the $\ct_i(0,t)$'s do not go to 0 as \teq{t \to \infty}.  The Froissart-Martin unitarity bounds \cite{froissart} in principle allow for asymptotic growth as fast as \teq{| \ct_{i}(0,t) | \sim \log^2 t}, which would require the dispersion relation in Eq. (\ref{disp}) to be once subtracted, introducing an undetermined, possibly negative, constant into the analytic expression for the \teq{\ct_i(0,0)}'s.\footnote{In \cite{chiralpositivity,ANSAR,jacques}, the Froissart-Martin bound was sufficient to ensure that the dispersion relation used to derive the analytic sign constraint did not require a subtraction.  Interestingly, in that case the sign constraints were also associated with violations of classical causality \cite{ANSAR}, which does not appear to be the case here.}

In Regge theory the asymptotic behavior is given by
\beq
\left| \ct_i (s,t) \right| \sim t^{\alpha(s)-1}
\label{regge}
\eeq
for very large $t>0$, where the function $\alpha(s)$ describes the leading Regge trajectory, associated with narrow resonances of two fermions that may be exchanged in the $t$-channel process.\footnote{For a modern review of Regge theory in the context of QCD see, for instance, \cite{regge}.}  Since $\Real \alpha(s)$ is generally expected to be a decreasing function of $s>0$, we conclude that in the presence of a narrow fermion-fermion resonance of mass $m_B$ and spin $J$, our positivity constraints must apply as long as
\beq
J = \Real \alpha(m_B^2) \leq 0~.
\eeq
That is, according to the Regge model, a violation of our positivity constraints would require the presence of a narrow fermion-fermion resonance with non-zero spin.  In the case of the nucleon-nucleon interaction there is such a bound state: the deuteron.

Let us study this example in more detail.  Low-energy nucleons may be described by an isospin doublet, $N$, interacting via a simple non-relativistic lagrangian,
\beqa
\La &=& N^\dagger \left[ i \partial_0 + \vv \nabla^2 / (2m) \right] N \nn
&& - \frac 1 2 \left[ C_S \left( N^\dagger N \right)^2 + C_T \left( N^\dagger \vv \sigma N \right)^2 \right] + \ldots
\label{KSW}
\eeqa
where the ellipses stand for both relativistic corrections to the propagator and 4-nucleon operators involving derivatives.  In the $\overline{MS}$ scheme \teq{C_S = -\ct_1+ \ct_2} and \teq{C_T= - \ct_4 - \ct_5}.

The scattering length for the $^3S_1$ channel (nucleons in the triplet spin state with zero orbital angular momentum) is given in the $\overline{MS}$ scheme by
\beq
a_{^3S_1} = \frac{m}{4\pi} \left(C_S + C_T\right)~,
\label{deut}
\eeq
which is negative as long as \teq{\ct_1 + \ct_4 + \ct_5 > \ct_2}.  Since $\ct_2$ (c.f. $a_{2}$ in Eq. (\ref{effective})) corresponds to a repulsive interaction, this will always be the case if the long-range nucleon-nucleon interaction is attractive.  The scattering length in the $^1S_0$ channel (nucleons in the singlet spin state with zero orbital angular momentum) is given by
\beq
a_{^1S_0} = \frac{m}{4\pi} \left(C_S - 3C_T\right)~,
\label{din}
\eeq
which will be negative as long as \teq{\ct_1 > \ct_2 + 3 \ct_4 + 3 \ct_5}.\footnote{The measured value of \teq{|C_T | = |\ct_4 + \ct_5|} is small, an observation that has been explained as a consequence of the theory of (\ref{KSW}) having an additional Wigner spin-isospin symmetry when $C_T=0$ \cite{MSW}.}  Negative scattering lengths for attractive potentials indicate the absence of bound states.\footnote{There are conflicting sign conventions in the literature for scattering length in terms of the scattering amplitude.  Special care is called for because there are also conflicting sign conventions for the scattering amplitude in terms of the $S$-matrix expectation value.}

\begin{figure}[t]

\includegraphics[scale=0.4]{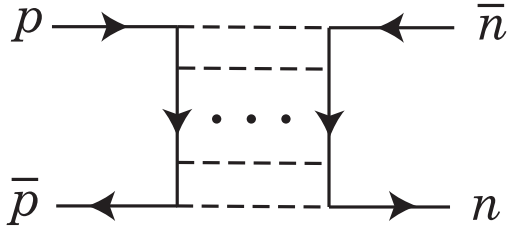}

\caption{Non-perturbative contribution to the spectral function in Eq. (\ref{disp}) that persists as $t \to \infty$ for small fixed $s$.  The solid lines represent nucleons (protons and neutrons) and the dashed lines represent pions.}
\label{nonpert}
\end{figure}

The experimental fact that $a_{^3S_1} > 0$, corresponding to the existence of the deuteron, implies that the dispersion relations in Eq. (\ref{disp}) require subtraction.  Since the deuteron is the only two-nucleon composite state with non-zero spin, the deuteron itself must lie on the leading Regge trajectory, preventing the spectral function \teq{\Imag c_1(0,t)} in Eq. (\ref{disp}) from vanishing for asymptotically large $t > 0$.  This is due to the non-perturbative contribution schematically represented in Fig. \ref{nonpert}, which essentially corresponds to exchanging a deuteron in the $t$-channel process.  This will be missed by any dispersion relation analysis of the nuclear interaction in which the spectral function is computed perturbatively.

See Appendix \ref{amplitudes} for a review of the physical meaning of the scattering length and for an explanation of why the $\ct_i(0,0)$'s discussed in Sec. \ref{dispersion} are linearly related to the scattering lengths.  For a review of the behavior of the scattering length as a function of the strength of the potential interaction, and of its connection to the presence or absence of bound states, see Appendix \ref{well}.

\section{Bound states and delta function potentials}
\label{deltaV}

In the $s$-channel a bound state is manifested as a pole in the $2 \to 2$ scattering amplitude of the form
\beq
{\cal M} \sim - \frac{m^2}{s - m_B^2}
\label{mBpole}
\eeq
where $m$ is the mass of the scattered particles and $m_B$ is the mass of the bound state.  This pole can only be seen non-perturbatively, by summing Feynman diagrams to all orders or by solving the Schr\"odinger equation.  At high energies the denominator in Eq. (\ref{mBpole}) gives the contact nucleon-nucleon interactions an anomalous dimension, driving them relevant (in the non-relativistic theory this corresponds to the KSW power-counting of \cite{KSW}).

\begin{figure}[t]

\includegraphics[scale=1.1]{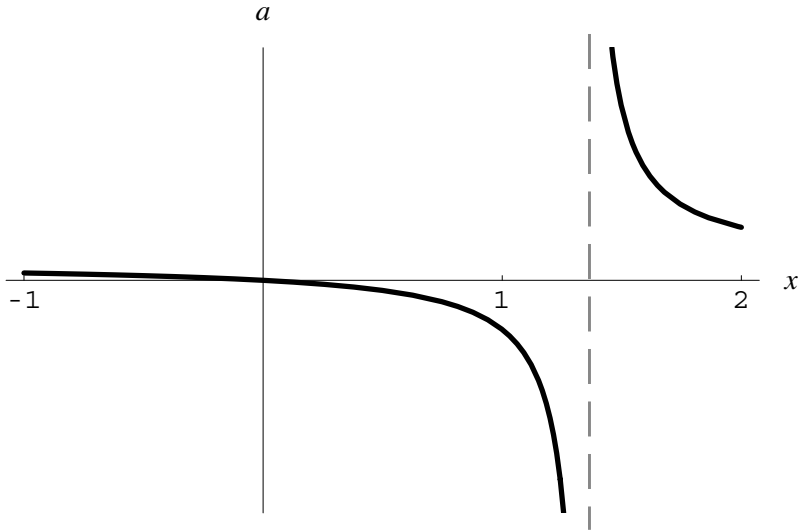}

\caption{Cartoon of the behavior of the scattering length $a$ as a function of $x \equiv g m / \mu$ for particles of mass $m$ interacting via the Yukawa potential given in Eq. (\ref{yukawaV}).}
\label{yukawa}
\end{figure}

In order to further illustrate the connection between the bound states, the asymptotic behavior of the scattering amplitude, and the sign of the effective low-energy interaction, let us consider the Yukawa interaction as a toy model. Non-relativistically this corresponds to a potential
\beq
V(r) = - \frac g r e^{-\mu r}~.
\label{yukawaV}
\eeq
Figure \ref{yukawa} gives a cartoon of the behavior of the scattering length $a$ as a function of \teq{x \equiv g m / \mu}.  A perturbative description is only appropriate for \teq{|x| \ll 1}.  For $x$ below about 1.35, the potential has no bound states.  As that critical value is approached from below, $a$ diverges, reflecting a change in the asymptotic behavior of the relevant spectral function.  For $x > 1.35$ the sign of $a$ has changed, reflecting the presence of bound states, and the dispersion integral must now require subtraction.

This divergence of the scattering length $a$ (and consequently of the zero-momentum total scattering cross-section, $\sigma_{\sl{tot}} = 4 \pi a^2$) as the critical value of $x$ is approached from above or from below, is an instance of the effect known as ``Sommerfeld enhancement'' \cite{sommerfeld}.  The Regge theory argument of Sec. \ref{asymptotic} suggests that this behavior can be embedded in a relativistic theory of fermions only if a bound state with non-zero spin develops as the coupling crosses the critical value.\footnote{The Bardeen-Cooper-Schrieffer (BCS) to Bose-Einstein condensate (BEC) crossover in ultracold atomic gases is another well-known example of a coupling passing through a critical value for which the scattering length $a$ diverges and changes sign \cite{BCS-BEC}.  At the critical value of the coupling, the atoms in the gas are said to be ``at unitarity'' \cite{bertsch}.}

Now consider an effective description of the physics at length scales much larger than $1/ \mu$.  We may then neglect the momentum dependence of the scattering, and the interaction potential effectively becomes a delta function in position,
\beq
V_{\sl{eff}}(r) = - \frac g {\mu^2} \delta^{(3)} (\vv r)~.
\eeq
This is the form of the potential which would arise from a non-relativistic theory with 4-fermion contact operators, such as Eq. (\ref{KSW}).  While classically irrelevant in three dimensions, strong renormalization can drive the delta function interaction relevant if the particle momenta are sufficiently large with respect to the scattering length \cite{luke}, i.e., for
\beq
| \vv p | \gtrsim 1 / | a |~.
\label{velocity}
\eeq
If $| a |$ were of order the inverse cutoff, \teq{\sim 1/ \mu}, then the interaction would be irrelevant in the domain of the effective theory.\footnote{For enlightening reviews of the delta function potentials in quantum mechanics and of the role of renormalization, see \cite{ajp,jackiw}.}  If $x$ is close to the critical value, however, then \teq{| a | \gg 1 / \mu}, which may be technically natural due to the presence of a non-relativistic conformal symmetry for a theory such as Eq. (\ref{KSW}) in the so-called ``unitarity limit,'' where \teq{|a_{^1S_0}|} and \teq{|a_{^3S_1}|} tend to infinity \cite{conformal}.  If the delta function potential is relevant, then for $a> 0$ it will have an $S$-wave bound state with binding energy \teq{E_B = 1 / 2 m a^2}.

\section{Quark mass dependence of the nuclear interaction}
\label{ineffective}

In recent years there has been considerable interest in understanding the dependence of the strong nuclear interaction at low energies on the mass of the pion, $m_\pi$ (or, equivalently, on the masses of the light quarks).  Numerical work on solving the Schr\"odinger equation with approximate potentials has yielded interesting but somewhat inconclusive results \cite{martin,epelbaum}.  Another approach, proposed by Donoghue \cite{donoghue} and recently further explored by Damour and Donoghue \cite{damour}, is to use the fixed-$s$ dispersion relation that we described in Sec. \ref{dispersion}, taking \teq{\Imag c_1(s,t)} from chiral perturbation theory supplemented by the Omn\`es function. 

One might naturally expect that the strength of the nuclear interaction would be controlled by $1/m_\pi^2$.  This would suggest that the effective 4-nucleon couplings $C_{S,T}$ in Eq. (\ref{KSW}) would be $ \sim 1/ m_\pi^2$, but this expectation fails by more than an order of magnitude, due to the largeness of \teq{|a_{^1S_0}|} and \teq{|a_{^3S_1}|}, which is a direct consequence of the fact that the deuteron has a small binding energy while the dineutron is very nearly bound.  Our work suggests that an understanding of the dependence of the nuclear interaction on $m_\pi$ must take into account the non-perturbative physics associated with the deuteron and dineutron.

This seems to us to be a case in which the naive expectation from effective field theory, that the strength of the low-energy interaction is controlled by the nearest singularity (pole or branch point) in the amplitude, fails to account for what we know non-perturbatively from the Schr\"odinger equation.\footnote{For another, more subtle instance of such ``ineffective field theories'' for the strong nuclear interaction, see \cite{bob}.}  For the nuclear interaction, Regge theory suggests that the presence of the deuteron bound state will change the asymptotic behavior of the fixed-$s$ dispersion integrals.  The values of $C_{S,T}$ are thus {\it not} controlled by the nearest singularity in the unsubtracted dispersion integral.  That the long-range nucleon-nucleon attraction might not be dramatically reduced by an increase in $m_\pi$ above its physical value also seems supported by results that suggest that the $f_0(600)$ resonance (also known as the $\sigma$) ---which is believed to account for most of the long-range nuclear attraction--- has a large admixture of a pion-pion bound state that persists as the pion mass is taken to be a few times its physical value \cite{boblattice,pelaez}.

\section{Weyl fermions}
\label{weyl}

In a theory without parity conservation and time-translation invariance, we must consider more invariant amplitudes than the five given by Eq. (\ref{fermionC}), but the signs of the low-energy invariant amplitudes considered in Sec. \ref{dispersion} should still be constrained as long as the relevant dispersion integrals converge.  It might therefore seem puzzling that those sign constraints are broken in a perturbative theory of Weyl fermions interacting via a complex scalar.\footnote{We thank Clifford Cheung for pointing out this theory to us.}  As we shall see, the problem here is that such a theory does not have the Regge behavior of Eq. (\ref{regge}), and the fixed-$s$ dispersion relation therefore needs to be subtracted even though there are no higher-spin bound states.

In this discussion we will use the notation of \cite{terning}, since it is the most consistent with the conventions we have used for Dirac fermions.  Consider a single Weyl fermion $\chi$, coupled to a complex scalar field $\Phi$:
\beq
{\cal L} =  i \chi^\dagger \bar \sigma^\mu \partial_\mu \chi - \frac m 2 (\chi \chi + \chi^\dagger \chi^\dagger) + (\partial_\mu \Phi^\ast) (\partial^\mu \Phi) - M^2 \Phi^\ast \Phi + g \Phi \chi \chi + g^\ast \Phi^\ast \chi^\dagger \chi^\dagger~.
\eeq
If \teq{M \gg m}, then when we integrate out $\Phi$ at energy scales far below $M$ the resulting effective action is
\beq
{\cal L}_{\sl{eff}} =  i \chi^\dagger \bar \sigma^\mu \partial_\mu \chi - \frac m 2 (\chi \chi + \chi^\dagger \chi^\dagger) + \frac{|g|^2}{M^2} \chi \chi \chi^\dagger \chi^\dagger~.
\label{badweyl}
\eeq
By the Fierz identities, we may rewrite the 4-fermion operator as
\beq
\chi \chi \chi^\dagger \chi^\dagger =  \frac 1 2 (\chi \sigma^\mu \chi^\dagger) (\chi \sigma_\mu \chi^\dagger)~.
\eeq
In terms of Dirac fermions we have that
\beq
\chi \sigma^\mu \chi^\dagger = \frac 1 2 \bar \psi \gamma^\mu \left( 1 + \gamma_5 \right) \psi~,
\eeq
so that the operator \teq{ (\chi \sigma^\mu \chi^\dagger) (\chi \sigma_\mu \chi^\dagger)} contains 
\beq
\frac 1 4 \left [ (\bar \psi \gamma^\mu \psi)^2 + (\bar \psi \gamma^\mu \gamma_5 \psi)^2 \right]~,
\eeq
which corresponds to an eigenvalue $-1$ of the matrix $F$ in Eq. (\ref{fierz}) and therefore is invariant under fierzing (see Sec. \ref{zeroenergy}).  The fact that the contact operator in Eq. (\ref{badweyl}) has a positive coefficient then conflicts with the analyticity constraints derived in Sec. \ref{dispersion}.

In the case of Dirac fermions, it was impossible for two particles (rather than a particle and an antiparticle) to annihilate into a boson, and at tree level there was therefore no pole in $s$ for the invariant amplitudes \teq{c_i(s,t)}.  In the case of Eq. (\ref{badweyl}), however, two Weyl fermions {\it can} annhiliate into a scalar, and the relevant invariant amplitude at tree level contains a term of the form
\beq
\frac 1 {s - M^2 + i \epsilon}~.
\label{spole}
\eeq
At fixed $s=0$, there must then be a contribution to $D^{(t)}(s,t',u')$ and $D^{(u)}(s,t',u')$ in Eq. (\ref{dispfull}) that does not vanish as we take $t' \to \infty$.  Therefore the integral cannot converge, so the dispersion relation requires subtraction.

In other words, the theory Eq. (\ref{badweyl}) will not exhibit the sort of Regge behavior described by Eq. (\ref{regge}), because the invariant \teq{c_i(s,t)} at fixed $s$ and asymptotically large $t$ is dominated by the tree-level contribution of Eq. (\ref{spole}), so that the Regge-theoretic argument for narrow resonances in the cross channel does not apply.  However, by an argument akin to that of \cite{jenkins}, we expect that the theory Eq. (\ref{badweyl}) will have a Lorentz-noninvariant vacuum with
\beq
\langle \chi \sigma^\mu \chi^\dagger \rangle \neq 0~,
\eeq
due to the particle-particle attraction, which makes it energetically favorable to form a condensate with non-zero fermion number.  Such a theory can thus break our sign constrains without having higher-spin bound states, but only at the cost of another IR-modification: a fermion condensate.  As in the case with bound states in the cross channel, this violation of the naive sign constraint implies that the quartic interaction is not, in fact, irrelevant.

\section{Discussion}
\label{discussion}
As we have seen, analyticity and unitarity enforce bounds on low-energy fermion scattering amplitudes through the analytic structure of the $S$-matrix.  For fermions these bounds translate into constraints on the leading 4-fermion contact interactions in an effective theory that admits an analytic UV-completion.  Those bounds may appear to be violated if the operators are driven relevant in the IR, as happens in the presence of poles from low-energy bound states.  In Regge theory, these bound states affect the asymptotic behavior of the spectral function that was used to deduce the sign constraints.  If a subtraction is needed in the dispersion relation, the constraints no longer apply.

In the case of the low-energy interaction between nucleons, we see that our positivity constraints do not apply because of the presence of a deuteron (spin-1) bound state which, through Regge theory, leads to the relevant dispersion requiring subtraction.  This suggests that the existence of the deuteron, which is a non-perturbative effect, must play a central role in any description of the strong nuclear force at low energies.  In a dispersion relation analysis it does so through its effect on the asymptotic behavior of the spectral function, and this indicates that the naive expectation from effective field theory, that the low-energy behavior is controlled by the nearest singularity in the scattering amplitude, fails for the strong nuclear interaction.

This also implies that analyticity alone forbids the existence of a dineutron (spin-0) bound state,\footnote{The reason for the name ``dineutron'' is that if the nucleons are in the spin singlet state with no orbital angular momentum, Fermi statistics require symmetrization in isospin.  Therefore both a proton-neutron and a dineutron bound state would have to exist.  (The diproton might be destroyed by the electromagnetic interaction.)} unless a deuteron bound state is also present.  This is so because in Regge theory a spin-0 state cannot change the asymptotic behavior of the spectral function in the fixed-$s$ dispersion relation so as to require that it be subtracted.  This observation might be useful in understanding the form of the dependence of the low-energy nuclear physics on $m_\pi$.  Notice that this result is consistent with the fact that lattice QCD simulations with varying pion masses all find that the  long-range attraction is stronger in the deuteron ($^3S_1$) than in the dineutron ($^1S_0$) channel \cite{aoki}.

In the case of Weyl fermions, we have written down a theory that does not exhibit the usual Regge behavior and therefore evades our sign constraints without resonances in the cross channel.  In this case, evasion of the constraint can be traced to a different IR modification of the effective field theory, the spontaneous breaking of Lorentz invariance by a condensate with non-zero fermion number.

One particularly interesting application of positivity constraints such as those we have considered is to the current-current interactions in the electroweak chiral lagrangian, where the associated dimension-6 operators are manifestly irrelevant \cite{buchmuller}, with no bound states or condensates to destroy the convergence of the dispersion integrals.  Fixing the signs of these operators is likely to have important consequences for precision electroweak constraints which both bound and depend on the values of these couplings (see, for instance, \cite{witek}).

It would be interesting to study these constraints in detail, but this would require us to generalize the formalism of Secs. \ref{invariants} and \ref{dispersion} to accomodate violations of parity and time-translation invariance.  More pressingly, we would also need to clarify whether non-Regge behavior such as as that identified for Weyl fermions in Eq. (\ref{badweyl}) can be avoided on general grounds, for example by forbidding spontaneous Lorentz violation.  Such matters are left for future research, but we conjecture that, as first suggested in \cite{O'Jenkins}, signs which in a narrow-resonance approximation require the presence of tachyons in the UV completion will turn out to be forbidden.

\appendix
\section{Invariant amplitudes and scattering lengths}
\label{amplitudes}

For a non-relativistic scattering process, the scattering length $a$ is defined by
\beq
\lim_{k \to 0} k \cot \delta_0 = - \frac{1}{a}~,
\eeq
where $k$ is the momentum of the scattered particles in the center-of-mass frame, and $\delta_0$ is the zeroth phase shift.  From this definition it follows that the total cross section for zero-momentum scattering is \teq{\sigma_{\sl{tot}} = 4 \pi a^2} and that
\beq
a \propto \frac{1}{m} {\cal M} (s = 4m^2, t=0, u=0)~.
\eeq
In the case of fermions, ${\cal M}$ may be written in terms of invariant amplitudes as shown in Eq. (\ref{fermionM}).  For zero momentum, the spinor products are all proportional to $m^2$, and therefore $a/m$ is a linear combination of the invariant \teq{c_i (4m^2, 0)}'s.

In the absence of a bound state or some other non-perturbative effect in the $s$-channel, the dependence of the $c_i$'s on $s$ between $s=0$ and $s=4m^2$ is likely to be mild, and therefore the appropriate linear combination of the $c_i(0,0)$'s would be a good approximation to the fermion-fermion scattering lengths.  But even if there is non-perturbative physics in the $s$-channel, we must remember that $c_i(0,0)$ is shorthand for
\beq
c_i(0,0,4m^2) = \ct_i(0,0,4m^2) - \ct_j(0,4m^2,0) F_{ij}
\eeq
and that when $m^2 \gg m_{\sl{gap}}^2$ (which is necessary to define a non-relativistic limit) the exchange term can be neglected, as we argued in Sec. \ref{dispersion}, so that
\beq
c_i(0,0,4m^2) \approx \ct_i (0,0,4m^2) = \ct_i (4m^2,0,0) \approx c_i (4m^2,0,0)~,
\eeq
where the second step follows from the so-called {\it crossing invariance}: any Feynman diagram for forward fermion-fermion scattering can be crossed into a diagram for forward fermion-antifermion scattering, and these will be equivalent up to the product of the external spinors, which is factored out to build the invariant amplitudes.  The contributions to any $\ct_i$ from the crossed diagrams are simply related by $s \leftrightarrow u$.\footnote{The crossing invariance can also been seen as a change of variables of integration in an action such as that of Eq. (\ref{Gamma1PI}).}

See also \cite{jackiw} for an argument why the renormalized coefficient of the delta-function potential (which should correspond to the appropriate linear combination of the renormalized $C_{S,T}$ couplings) is proportional to the corresponding scattering lengths.

\section{Scattering length for a spherical well}
\label{well}

Consider a spherical well potential in three dimensions, with range $L$ and depth $V_0$.  In terms of \teq{u(r) \equiv r \psi (r)}, Schr\"odinger's equation for the mode with vanishing energy and angular momentum is just
\beq
u''(r) \propto V(r) \times u(r)~,
\eeq
i.e., $u$ is oscillatory inside the square well (where \teq{V(r) = - V_0}) and linear outside it (where \teq{V(r) = 0}).  The scattering length $a$ corresponds to the $r$-intercept of this line.

\begin{figure}
\centering
\subfigure[]{\includegraphics[width=0.3\textwidth]{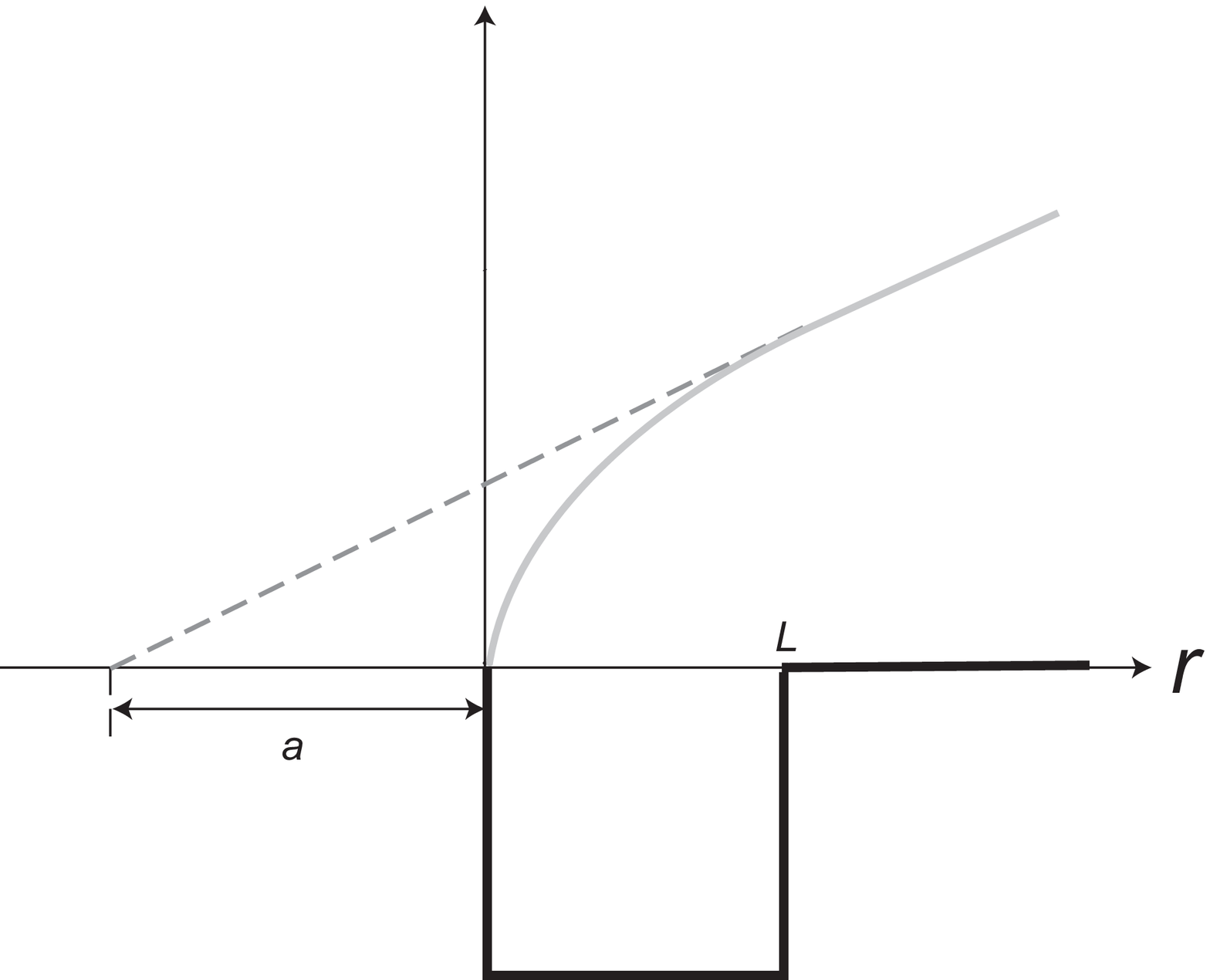}}\\
\subfigure[]{\includegraphics[width=0.22\textwidth]{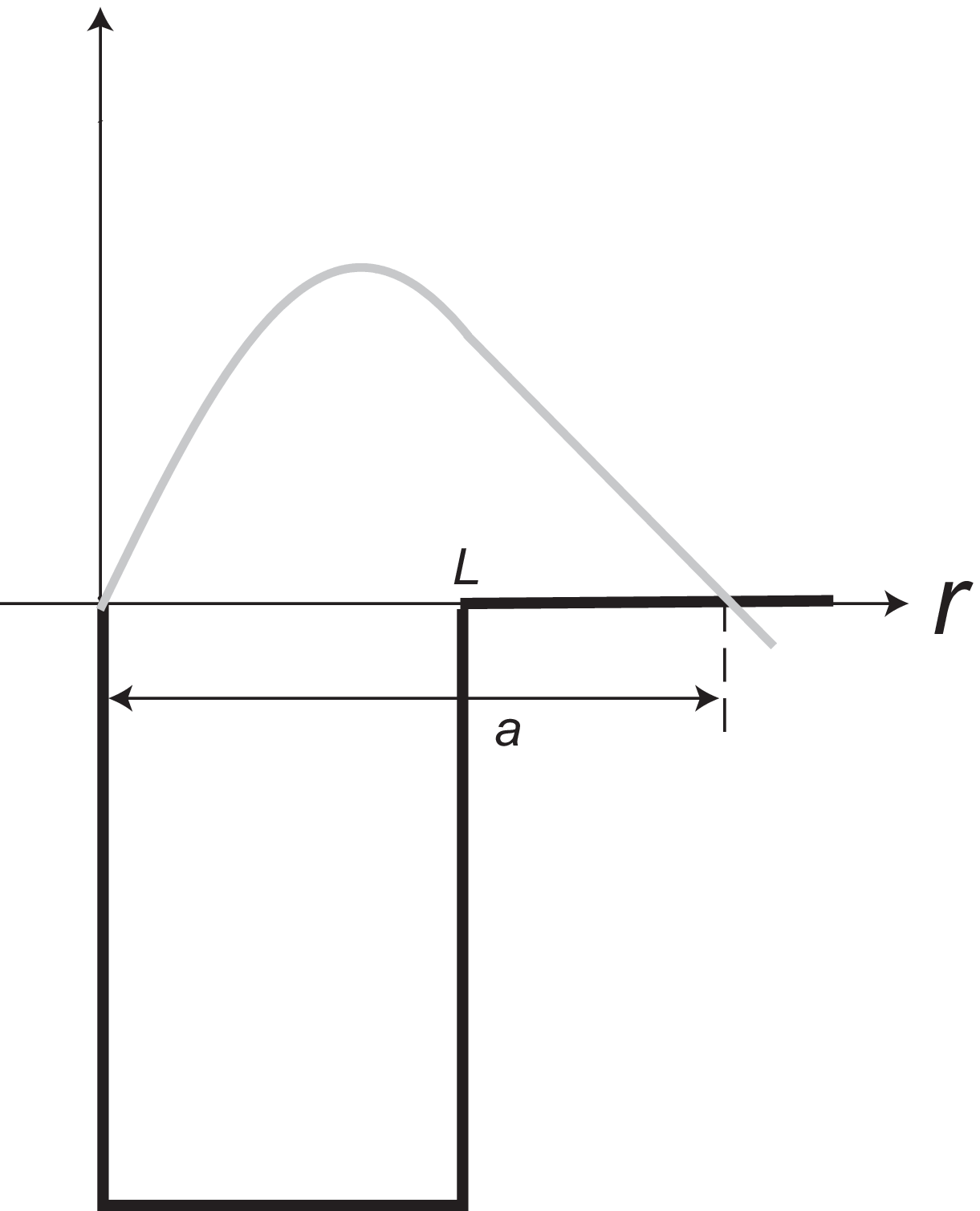}}\hspace{2cm}
\subfigure[]{\includegraphics[width=0.22\textwidth]{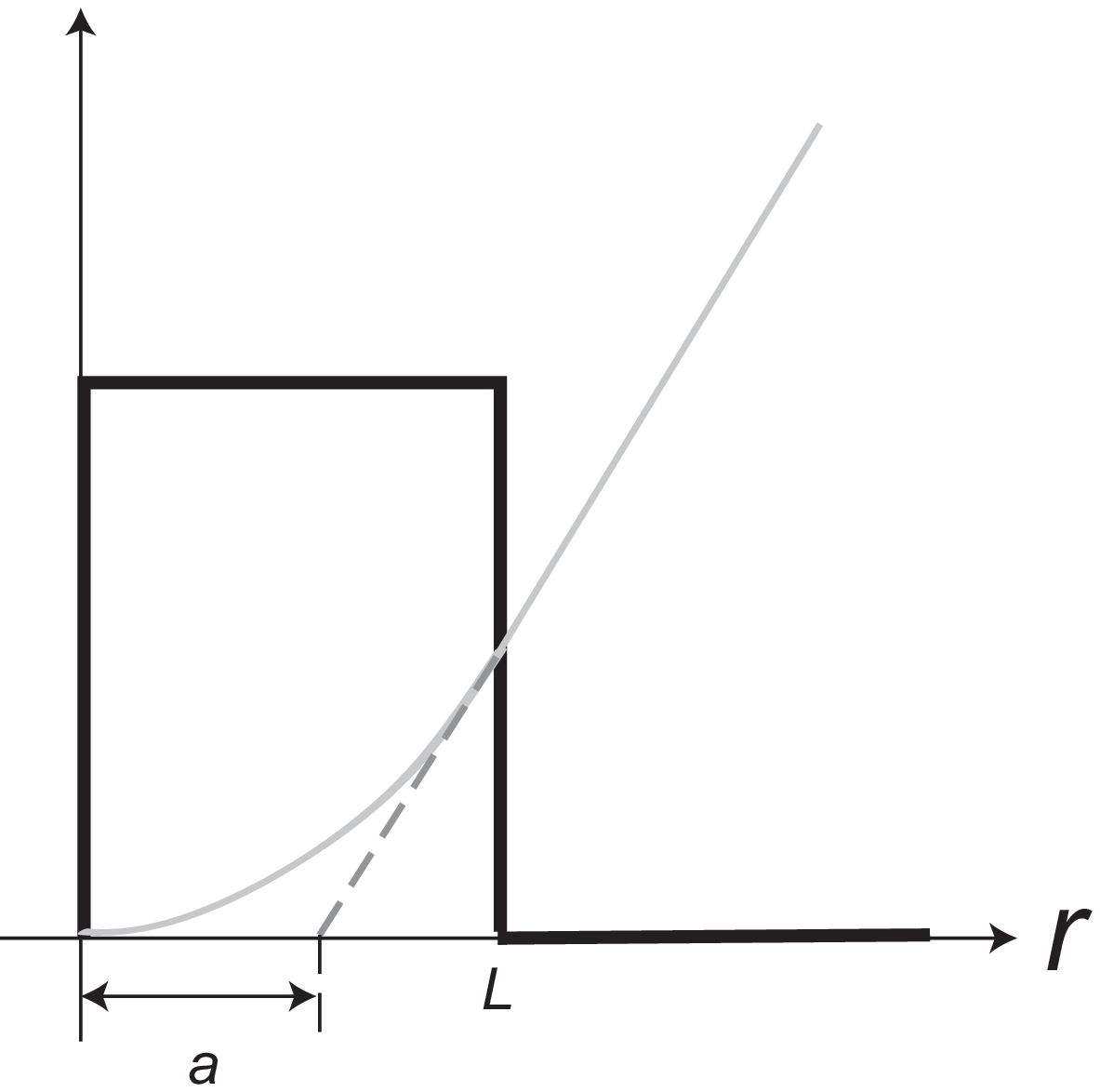}}
\caption{Plots of $u(r) \equiv r \psi (r)$ (in gray) for solutions to the Schr\"odinger equation in three dimensions with $E = l =0$, for a square well potential (plotted in black) with depth $V_0$ and range $L$.  Only $V_0$ varies between the plots shown in (a), (b), and (c).  The value of the $r$-intercept is equal to the scattering length $a$.\label{square}}
\end{figure}

For a shallow attractive potential, as shown in Fig. \ref{square}(a), the scattering length $a$ is strictly negative.  For a sufficiently deep well, as shown in Fig. \ref{square}(b), the wavefunction will have a zero at $a > L > 0$ (i.e., the scattering length will be positive and greater than the range of the potential).  For negative $V_0$, as shown in Fig. \ref{square}(c), the potential is repulsive and the intercept of $u(r)$ is positive, but, unlike the case of the attractive potential, the scattering length is strictly less than the classical range of the potential, \teq{0 < a < L}.

Note what this means about the behavior of the scattering length as we vary the depth of the well:  For negative $V_0$ the potential is repulsive.  There is no bound state and $a$ is negative.  As $V_0$ goes to zero, both the interaction and the scattering length vanish.  For small $V_0 > 0$, the scattering length is negative and there is no bound state.  As we make the well deeper, the wavefunction will curve over more and more inside the well, and the intercept $a$ will grow more and more negative.  Eventually we approach a critical depth $V_0^c$, where $a$ runs to $- \infty$  If we make the well any deeper, the wavefunction turns over inside the well, leading to a large positive scattering length.  As we make the well deeper, the scattering length decreases, asymptotically approaching the classical range of the potential, $L$.

This example illustrates a general feature of the behavior of scattering lengths:  when we vary some parameter of a central potential such that a bound state appears or disappears, the corresponding scattering length passes through infinity and changes sign.  By contrast, the scattering length passing through zero does not indicate the (dis)appearance of a bound state, but rather corresponds to the vanishing of the interaction.

\begin{acknowledgments}
We thank Paulo Bedaque for suggesting the issue of deuteron binding and John Donoghue for introducing us to the literature on fixed-$s$ dispersion relations for nucleon scattering, as well as for his comments and questions on the manuscript.  We also thank Bob Jaffe for pointing out an error in our understanding of low-energy scattering, and Nima Arkani-Hamed, Cliff Cheung, Jacques Distler, John McGreevy, Tom Mehen, Michael Peskin, Iain Stewart, and Mark Wise for discussions.  This work was supported in part by the U.S. Department of Energy under contracts DE-FG03-92ER40701 (A.A. and A.J.), DE-FG02-05ER41360, and DE-FG02-90ER40542 (D.O.).
\end{acknowledgments}

\end{document}